%% file: emuchar_parco.tex
\newif\ifdraft
\ifdraft 
\documentclass[draft]{elsarticle}
\usepackage{lineno}
\linenumbers
\onecolumn
\else
\documentclass[5p,times]{elsarticle}
\fi
\usepackage{microtype}
\usepackage{booktabs} 
\usepackage{comment} 
\usepackage{paralist}
\usepackage{listings}
\usepackage{parcolumns}

\usepackage{amsmath}

\usepackage{graphicx}
\usepackage{subcaption}
\usepackage{adjustbox}

\usepackage{tikz}
\usetikzlibrary{shapes,shapes.geometric,arrows.meta,fit,calc,positioning,chains}

\title{A Microbenchmark Characterization of the Emu Chick}

\author[2]{Jeffrey S. Young\corref{cor1}}
\ead{jyoung9@gatech.edu}
\author[3]{Eric Hein}
\ead{ehein6@gatech.edu}
\author[1]{Srinivas Eswar}
\ead{seswar3@gatech.edu}
\author[1]{Patrick Lavin}
\ead{plavin3@gatech.edu}
\author[4]{Jiajia Li}
\ead{jiajiali@gatech.edu}
\author[1]{Jason Riedy}
\ead{jason.riedy@gatech.edu}
\author[1]{Richard Vuduc}
\ead{richie@gatech.edu}
\author[2]{Tom Conte}
\ead{conte@gatech.edu}

\address[1]{School of Computational Science and Engineering, Georgia Institute of Technology, North Avenue, Atlanta, GA 30332}
\address[2]{School of Computer Science, Georgia Institute of Technology, North Avenue, Atlanta, GA 30332}
\address[3]{Emu Technology, 270 West 39th Street, 13th Floor,
New York, NY 10018}
\address[4]{Pacific Northwest National Laboratory, 902 Battelle Blvd, Richland, WA 99354}

\cortext[cor1]{Corresponding author, jyoung9@gatech.edu}

\begin{document}
\thispagestyle{empty}


\begin{abstract}
\input{sections/abstract.tex}
\end{abstract}

\maketitle
 \thispagestyle{plain}
 \pagestyle{plain}

\input{sections/intro.tex}
\input{sections/background.tex}
\input{sections/exp_setup.tex}
\input{sections/results.tex}

\input{sections/discussion.tex}
\input{sections/related-work.tex}
\input{sections/conclusion.tex}

\section{Acknowledgments}\label{sec:acknowledgments}

This work was supported in parts by NSF Grant ACI-1339745 (XScala), NSF Grant OAC-1710371 (SuperSTARLU), an IARPA contract, and the Defense Advanced Research Projects Agency (DARPA) under agreement \#HR0011-13-2-0001. Any opinions, findings, conclusions, or recommendations in this paper are solely those of the authors and do not necessarily reflect the position or the policy of the sponsors. 

The authors also gratefully acknowledge support by the Laboratory Directed Research and Development program at Sandia National Laboratories, a multi-mission laboratory managed and operated by National Technology and Engineering Solutions of Sandia, LLC, a wholly owned subsidiary of Honeywell International, Inc., for the U.S. Department of Energy's National Nuclear Security Administration under contract DE-NA0003525. 

Finally, thanks to the Emu Technology team for their continued support and debugging assistance with the Emu Chick prototype and to the many reviewers with their helpful suggestions.

\bibliographystyle{elsarticle-num-names}
\bibliography{bib/refs} 

\end{document}

%% file: sections/abstract.tex
The Emu Chick is a prototype system designed around the concept of migratory memory-side processing. Rather than transferring large amounts of data across power-hungry, high-latency interconnects, the Emu Chick moves lightweight thread contexts to near-memory cores before the beginning of each memory read.
The current prototype hardware uses FPGAs to implement cache-less ``Gossamer'' cores for computational work and rely on a typical stationary core (PowerPC) to run basic operating system functions and migrate threads between nodes.
In this multi-node characterization of the Emu Chick, we extend an earlier single-node investigation \cite{hein:2018:ashes_emu} of the the memory bandwidth characteristics of the system through benchmarks like STREAM, pointer chasing, and sparse matrix-vector multiplication. We compare the Emu Chick hardware to architectural simulation and an Intel Xeon-based platform. Our results demonstrate that for many basic operations the Emu Chick can use available memory bandwidth more efficiently than a more traditional, cache-based architecture although bandwidth usage suffers for computationally intensive workloads like SpMV. Moreover, the Emu Chick provides stable, predictable performance with up to 65\% of the peak bandwidth utilization on a random-access pointer chasing benchmark with weak locality.


%% file: sections/intro.tex
\section{Introduction} \label{sec:intro}

Analysis of data represented as graphs, sparse tensors, and other non-regular structures poses many challenges for traditional computer architectures because the data locality of these applications typically occurs in small bursts. While individual data elements may have multiple associated attributes nearby (\textit{e.g.} neighbors, weights, semantic attributes, and timestamps for streaming graph edges), analysis algorithms tend to access these small chunks in a more random fashion. Limited spatial locality in traditional analysis kernels leads to underutilizing cache lines, confounding prefetch engines, and thus reducing overall effective memory bandwidth. Furthermore, common analysis kernels may exhibit dynamic parallelism and create many data-dependent memory references.  These references can stall architectures that cannot maintain enough contexts and requests in flight.
%
Consequently, today's ``big data''
platforms frequently are outperformed by a single thread
accessing a large SSD~\cite{189908}.

This state of affairs motivates novel architectures like the Emu migratory thread system~\cite{dysart2016emu}, the subject of this study.
The Emu is a cache-less system built around ``nodelets'' that each execute lightweight threads.  These threads migrate to data rather than moving data through a traditional cache hierarchy. 

This paper expands on the first independent characterization of the Emu Chick prototype\cite{hein:2018:ashes_emu} by exploring multiple distributed nodes that consist of those nodelets (see Section~\ref{sec:background}). Our study uses microbenchmarks and small kernels---namely, STREAM, pointer chasing, and sparse matrix-vector multiplication (SpMV)---as proxies that reflect some of the key characteristics of our motivating computations, which come from sparse and irregular applications~\cite{stinger-hpec12,li:2016:ttm}.
Indeed, one larger goal of our work beyond this paper is to develop a performance-portable, Emu-compatible API for 
Georgia Tech's STINGER open-source streaming graph framework~\cite{stinger-hpec12} and 
ParTI~\cite{parti:2018:gh} tensor decomposition algorithms (\textit{e.g.} CP and Tucker decomposition).
Mapping such applications to the Emu architecture is difficult because the thread migration makes programming around \emph{data's location} critical to reducing migrations.


This study's specific demonstrations include
\begin{itemize}
\item a detailed characterization of the Emu Chick hardware using custom Cilk kernels derived from optimized OpenMP kernels;
\item an analysis of memory bandwidth on the Chick system and comparison to a more traditional cache-based architecture with Emu results tested on 64 nodelets
across eight nodes
\item a discussion of memory allocation, data layout, and ``smart'' thread migration on the Emu architecture with respect to SpMV kernels;
\item an investigation and validation of the Emu architectural simulator for projecting larger configurations' performance. 
\end{itemize}
The main high-level finding is that an Emu-style architecture can more efficiently utilize available memory bandwidth while reducing the variability of that bandwidth to the memory access pattern. However, achieving such results still requires careful consideration of the interplay between data layout and its affect on thread migration. Additionally, our current Chick prototype is still compute-bound for some algorithms like SpMV which hurts its usage of available memory bandwidth when compared to a traditional CPU-based system.


%% file: sections/background.tex
\section{The Emu Architecture} \label{sec:background}

\begin{figure}
\centering 
\includegraphics[width=0.7\linewidth]{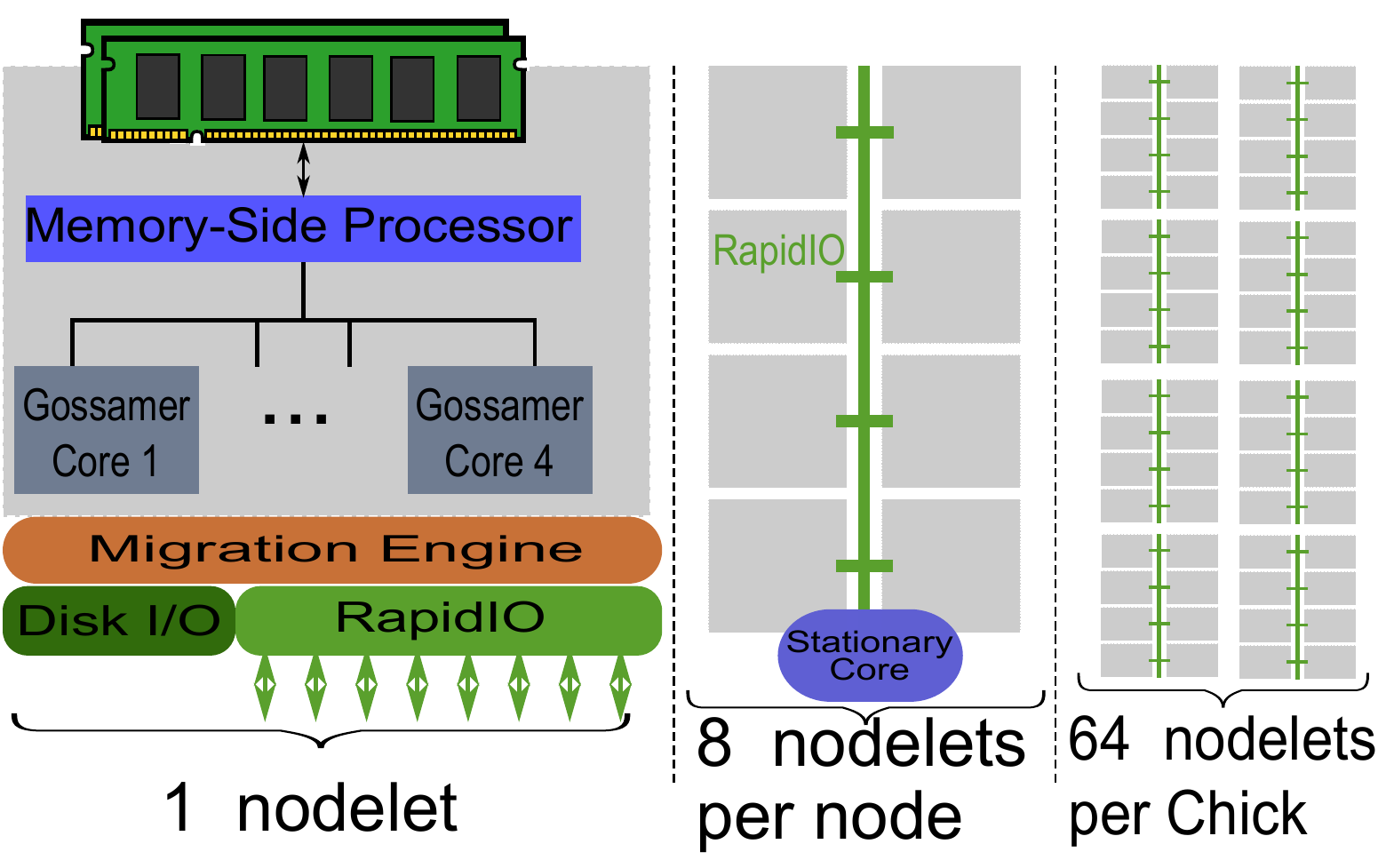}
\caption{Emu architecture: The system consists of \emph{stationary} processors for running the operating system and up to four \emph{Gossamer} processors per nodelet tightly coupled to memory.  The cache-less Gossamer processing cores are multi-threaded to both source sufficient memory references and also provide sufficient work with many outstanding references.  The coupled memory's narrow interface ensures high utilization for accesses smaller than typical cache lines.}\label{fig:emu-arch}
\end{figure}

The Emu architecture focuses on improved random-access bandwidth scalability by migrating lightweight, \emph{Gossamer} threads or ``threadlets'' to data and emphasizing fine-grained memory access.
A general Emu system consists of the following processing elements, as illustrated in Figure~\ref{fig:emu-arch}:
\begin{itemize}
\item A common \emph{stationary} processor runs the operating system (\textit{e.g.} Linux) and manages storage and network devices.
\item \emph{Nodelets} combine narrowly banked memory (NCDIMMs) with several highly multi-threaded, cache-less \emph{Gossamer} cores to provide a memory-centric environment for migrating threads.
\end{itemize}
These elements are combined into nodes that are connected by a RapidIO fabric. The current generation of Emu systems include one stationary processor for each of the eight nodelets contained within a node. System-level storage is provided by SSDs. We talk more specifically about some of the prototype limitations of our Emu Chick in Section~\ref{sec:exp}. A more detailed description of the Emu architecture is available elsewhere~\cite{dysart2016emu}.

For programmers, the Gossamer cores are transparent accelerators. The compiler infrastructure compiles the parallelized code for the Gossamer ISA, and the runtime infrastructure launches threads on the nodelets. Currently, one programs the Emu platform using Cilk~\cite{leiserson1997programming}, providing a path to running on the Emu for simple OpenMP programs whose translations to Cilk are straightforward.
The current compiler supports the expression of task or fork-join parallelism through Cilk's \texttt{cilk\_spawn} and \texttt{cilk\_sync} constructs, with a future Cilk Plus software release in progress that would include \texttt{cilk\_for} (the nearly direct analogue of OpenMP's \texttt{parallel for}). Many existing C and C++ OpenMP codes can translate almost directly to Cilk Plus.

A launched Gossamer thread only performs local reads. Any remote read triggers a migration, which will transfer the context of the reading thread to a processor local to the memory channel containing the data.  Experience on high-latency thread migration systems like Charm++ identifies migration overhead as a critical factor even in highly regular scientific codes~\cite{7013040}. The Emu system keeps thread migration overhead to a minimum by limiting the size of a thread context, implementing the transfer efficiently in hardware, and integrating migration throughout the architecture. In particular, a Gossamer thread consists of 16 general-purpose registers, a program counter, a stack counter, and status information, for a total size of less than 200 bytes.  The compiled executable is replicated across the cores to ensure that instruction access always is local. Limiting thread context size also reduces the cost of spawning new threads for dynamic data analysis workloads. Any operating system requests are forwarded to the stationary control processors through the service queue.

The highly multi-threaded Gossamer cores, which are reading only local memory, do not need caches nor, therefore, cache coherency traffic.  Additionally, ``memory-side processors'' provide atomic read or write operations that can be used to access small amounts of data without triggering unnecessary thread migrations. A node's memory size is relatively large (64\,GiB) but with multiple, narrow memory channels (8 channels with 8 bit interfaces), in order to extract weak spatial locality from data analysis kernels while maintaining low-latency read and write operations. The high degree of multi-threading also helps to cover the migration latency of the many threadlets.  The Emu architecture is designed from the ground up to support high bandwidth utilization and efficiency for demanding data analysis workloads.


%% file: sections/exp_setup.tex
\section{Experimental Setup} \label{sec:exp}

\subsection{Emu Chick Prototype}
\label{sec:prototype}

The Emu Chick prototype is still in active development. The current hardware iteration uses an Arria 10 FPGA on each node card to implement the Gossamer cores, the migration engine, and the stationary cores. Several aspects of the system are scaled down in the current prototype with respect to the next-generation system which will use larger and faster FPGAs to implement computation and thread migration. The current Emu Chick prototype has the following features and limitations:

\begin{itemize}
\item Our system has one Gossamer Core (GC) per nodelet with a concurrent max of 64 threadlets. The next-generation system will have four GC's per nodelet, supporting 256 threadlets per nodelet.
\item A full Chick system has 64 nodelets across eight nodes, implementing a distributed Partitioned Global Address System (PGAS) architecture that is connected by the RapidIO network.
\item Our GC's are clocked at 175MHz currently (up from 150MHz in~\cite{hein:2018:ashes_emu}) rather than the planned 300MHz for later-generation Emu development systems.
\item The Emu's DDR4 DRAM modules are clocked at 1600MHz rather than the full 2133MHz. 
\item The current Emu software version provides support for C++ but does not yet include functionality to translate Cilk Plus features like \texttt{cilk\_for} or Cilk \texttt{reducers}~\cite{Frigo:2009:ROC:1583991.1584017}. All benchmarks currently use \texttt{cilk\_spawn} directly, which also allows more control over spawning strategies.

\end{itemize}

All experiments are run using Emu's 18.09 compiler and simulator toolchain, and the Emu Chick system is running NCDIMM firmware version 2.5.1, system controller software version 3.1.0, and each stationary core is running the 2.2.3 version of software.

\subsection{Emu Simulator}
Emu provides a timing simulator implemented using SystemC, and this simulator can be used to test and evaluate software before running on the hardware. Previous characterization experiments in \cite{hein:2018:ashes_emu} employed a configuration of the simulator to match the characteristics of a single node (8 nodelets) of the Chick hardware for validation and to do basic projections to a stable, 64-node Chick system. We have not repeated these experiments as the projections in earlier work have been superseded by real-time results on the 64-node Chick system and many of the most interesting characterizations cannot be run on the timing simulator.

\subsection{Common CPU-Focused Comparison Platform}
\label{sec:cpu-specs}

In order to make an initial comparison of the Emu's \mbox{memory} bandwidth characteristics with commodity hardware, each benchmark is also run on an four-socket Intel Xeon E7-4850 v3  (Haswell) machine with 2 TiB of DDR4 (referred to as \texttt{Haswell Xeon} in associated results). The CPUs on the Haswell server are each clocked at 2.20GHz and each have a 35 MiB L3 cache, while the memory is clocked at 1333 MHz (although it is rated for 2133 MHz, the risers used to increase density decrease the frequency). Each socket has a peak theoretical bandwidth of 42.7 GB/s because of the underclocked memory.


For each benchmark, Emu-specific intrinsics (\textit{e.g.} localized mallocs) are swapped out for their x86 equivalents, and the benchmarks are compiled with GCC 5.5.0. The Cilk keywords are left unchanged, allowing GCC's Cilk runtime to implement the parallel functionality. Intel's MKL library (version 20180001) is used for some of the SpMV comparisons made in Section \ref{sec:spmv_results}. STREAM is run with default OpenMP settings including OMP\_PROC\_BIND=false and OMP\_SCHEDULE=static. 

\subsection{Metrics for Comparing the Emu Prototype with Cache-Based Hardware}
The architectural design choices that enable the Emu computational model (migrate threads instead of data, narrow memory channels, limited thread context) and the base platforms for the prototype (FPGAs with lower clock frequencies) make it difficult to accurately 
compare the Emu and CPU- or GPU-focused systems in terms of execution or runtime. 

Additionally, the Emu platform uses Narrow-Channel DRAM (NCDRAM) which reduces the width of the DRAM bus to 8 bits. Otherwise, the memory uses standard DDR4 chips. An 8-byte word can be transferred in a single burst. The smaller bus means that each channel of NCDRAM has only 2GB/s of bandwidth, but the system makes up for this by having many more independent channels. Because of this, it can sustain more simultaneous fine-grained accesses than a traditional system with fewer channels and the same peak memory bandwidth specification.

Due to difficulties in comparing differently clocked architectures with different memory controller configurations, we focus our initial characterization not on runtime but on memory bandwidth (MB/s) and effective memory bandwidth utilization (\% of measured peak memory bandwidth). In a CPU-focused system, this might be analogous to effective cache line utilization while in the Emu it correlates more closely to how much bandwidth can be achieved with respective to other system overheads, such as thread migration and queuing delays.

\subsection{Benchmarks}
                                                  \label{ssec:bmks}
As discussed in Section~\ref{sec:prototype}, the Emu Chick toolchain currently lacks support for \texttt{cilk\_for} and Cilk reducers. However, we present several benchmarks that use Cilk semantics to characterize the performance of the system, specifically focusing on kernels that expose the memory bandwidth characteristics of the system and test important kernels like SpMV that are key for applications like sparse tensor decomposition.  For each benchmark result, we present the average memory bandwidth (usually expressed as megabytes per second) over ten trials.

\paragraph{STREAM}
The STREAM benchmark~\cite{McCalpin1995} has been ported and tuned for the Emu hardware to measure raw memory bandwidth. The ADD kernel computes the vector sum of two large arrays of 8-byte integers, storing the result in a third array. On the Emu Chick these arrays are striped across all the nodelets in the system.

This benchmark demonstrates that thread spawning is important for performance.  Different thread spawning mechanisms achieve different memory bandwidths.
Our tests control thread spawning and do not rely on \texttt{cilk\_for}.
The spawning methods follow trees that are briefly described as follows:
\begin{itemize}
\item \textbf{serial\_spawn}: threads spawn locally on a single nodelet using a for loop,
\item \textbf{recursive\_spawn}: threads are spawned locally using recursive calls,
\item \textbf{serial\_remote\_spawn}: threads are spawned on each nodelet, which in turn uses a for loop to spawn threads locally, and
\item \textbf{recursive\_remote\_spawn}: threads are spawned recursively across all nodelets, and then each nodelet recursively spawns new threads locally.
\end{itemize}


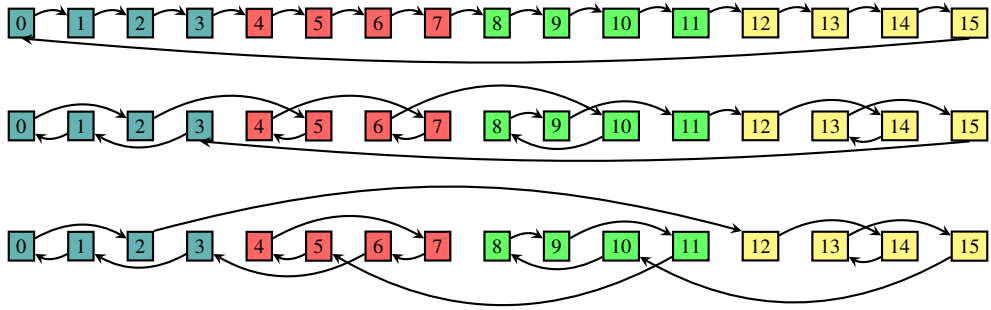
\begin{figure*}\centering

  \begin{tikzpicture}[start chain=going right, node distance=5mm,>=stealth,thick,font=\small,scale=0.85, every node/.style={transform shape}]
    \foreach \i in {0,...,3} \node (n\i) [draw,on chain,fill=teal!60] {\i};
    \foreach \i in {4,...,7} \node (n\i) [draw,on chain,fill=red!60] {\i};
    \foreach \i in {8,...,11} \node (n\i) [draw,on chain,fill=green!60] {\i};
    \foreach \i in {12,...,15} \node (n\i) [draw,on chain,fill=yellow!60] {\i};

    \def\topfrom{0,1,2,3,4,5,6,7,8,9,10,11,12,13,14}
    \def\topto{1,2,3,4,5,6,7,8,9,10,11,12,13,14,15}

    \def\exceptionfrom{15}
    \def\exceptionto{0}

    \foreach \from [count=\c,evaluate=\c as \to using {{\topto}[\c-1]}]  in \topfrom
      \draw [->] (n\from) to [bend left=30] (n\to);

    \foreach \from [count=\c,evaluate=\c as \to using {{\exceptionto}[\c-1]}]  in \exceptionfrom
      \draw [->] (n\from.south) to [bend left=5] (n\to.south);
  \end{tikzpicture}
  
  \begin{tikzpicture}[start chain=going right, node distance=5mm,>=stealth,thick,font=\small,scale=0.85, every node/.style={transform shape}]
    \foreach \i in {0,...,3} \node (n\i) [draw,on chain,fill=teal!60] {\i};
    \foreach \i in {4,...,7} \node (n\i) [draw,on chain,fill=red!60] {\i};
    \foreach \i in {8,...,11} \node (n\i) [draw,on chain,fill=green!60] {\i};
    \foreach \i in {12,...,15} \node (n\i) [draw,on chain,fill=yellow!60] {\i};

    \def\topfrom{3,1,0,2, 5,4,7,6, 10,8,9,11, 12,14,13} 
    \def\topto{1,0,2, 5,4,7,6, 10,8,9,11, 12,14,13,15} 

    \def\exceptionfrom{15}
    \def\exceptionto{3}

    \foreach \from [count=\c,evaluate=\c as \to using {{\topto}[\c-1]}]  in \topfrom
      \draw [->] (n\from) to [bend left=30] (n\to);

    \foreach \from [count=\c,evaluate=\c as \to using {{\exceptionto}[\c-1]}]  in \exceptionfrom
      \draw [->] (n\from.south) to [bend left=5] (n\to.south);
  \end{tikzpicture}

  \begin{tikzpicture}[start chain=going right, node distance=5mm,>=stealth,thick,font=\small,scale=0.85, every node/.style={transform shape}]
    \foreach \i in {0,...,3} \node (n\i) [draw,on chain,fill=teal!60] {\i};
    \foreach \i in {4,...,7} \node (n\i) [draw,on chain,fill=red!60] {\i};
    \foreach \i in {8,...,11} \node (n\i) [draw,on chain,fill=green!60] {\i};
    \foreach \i in {12,...,15} \node (n\i) [draw,on chain,fill=yellow!60] {\i};

    \def\topfrom{5,4,7,6, 3,1,0, 12,14,13,15, 10,8,9,11} 
    \def\topto{4,7,6, 3,1,0,2, 14,13,15, 10,8,9,11, 5}  

    \def\exceptionfrom{2}
    \def\exceptionto{12}

    \foreach \from [count=\c,evaluate=\c as \to using {{\topto}[\c-1]}]  in \topfrom
      \draw [->] (n\from) to [bend left=30] (n\to);

    \draw [->] (n2.north east) to [bend left=15] (n12.north west);
  \end{tikzpicture}

  \caption{(top) An ordered linked list, in which consecutive elements have sequential memory addresses, (middle) A linked list with a intra-block shuffle permutation applied to randomize the ordering of elements within a block. Note that all elements within a block are accessed before jumping to the next block. (bottom) A linked list with a full block shuffle permutation applied. Not only are the elements within a block shuffled, but the traversal order of the blocks themselves has also been randomized.  Not shown is the full shuffle which is equivalent to a block shuffle followed by an intra-block shuffle. }
  \label{fig:permutations}
\end{figure*}

\paragraph{Pointer Chasing}
In this benchmark, each thread adds up all the elements in a linked list. Each element consists of an 8-byte payload and an 8-byte pointer to the next element. After the elements of this linked list are grouped into blocks, their ordering is randomized. This permutation may be applied to the ordering of the elements within each block (\texttt{intra\_block\_shuffle}), or the ordering of the blocks themselves (\texttt{block\_shuffle}), or both (\texttt{full\_block\_shuffle}). The block size is also varied to emulate different levels of spatial locality that may arise in a workload. Figure \ref{fig:permutations} explains the list initialization further. 

The pointer chasing benchmark has three key properties by design. 
\begin{itemize}
\item Data-dependent loads: Memory-level parallelism is severely limited since each thread must wait for one pointer dereference to complete before accessing the next pointer
\item Fine-grained accesses: Spatial locality is restricted since all accesses are at a 16B granularity. This is smaller than a 64B cache line on x86 platforms, and much smaller than a typical DRAM page size.
\item Random access pattern: Since each block of memory is read exactly once in random order, caching and prefetching are mostly ineffective.
\end{itemize}

The pointer chasing benchmark simulates a worst-case memory fragmentation scenario that can arise in memory intensive workloads such as streaming graph analytics. When small list elements are dynamically allocated and deallocated from a shared memory pool, the resulting data structure will exhibit all of these characteristics when it is traversed. The pointer chase benchmark otherwise is quite similar to the GUPS/RandomAccess benchmark\cite{Luszczek_2006}, however GUPS lacks data-dependent loads, and pointer chase does not modify the list. 

\begin{figure}
  \centering
  \includegraphics[width=0.8\linewidth]{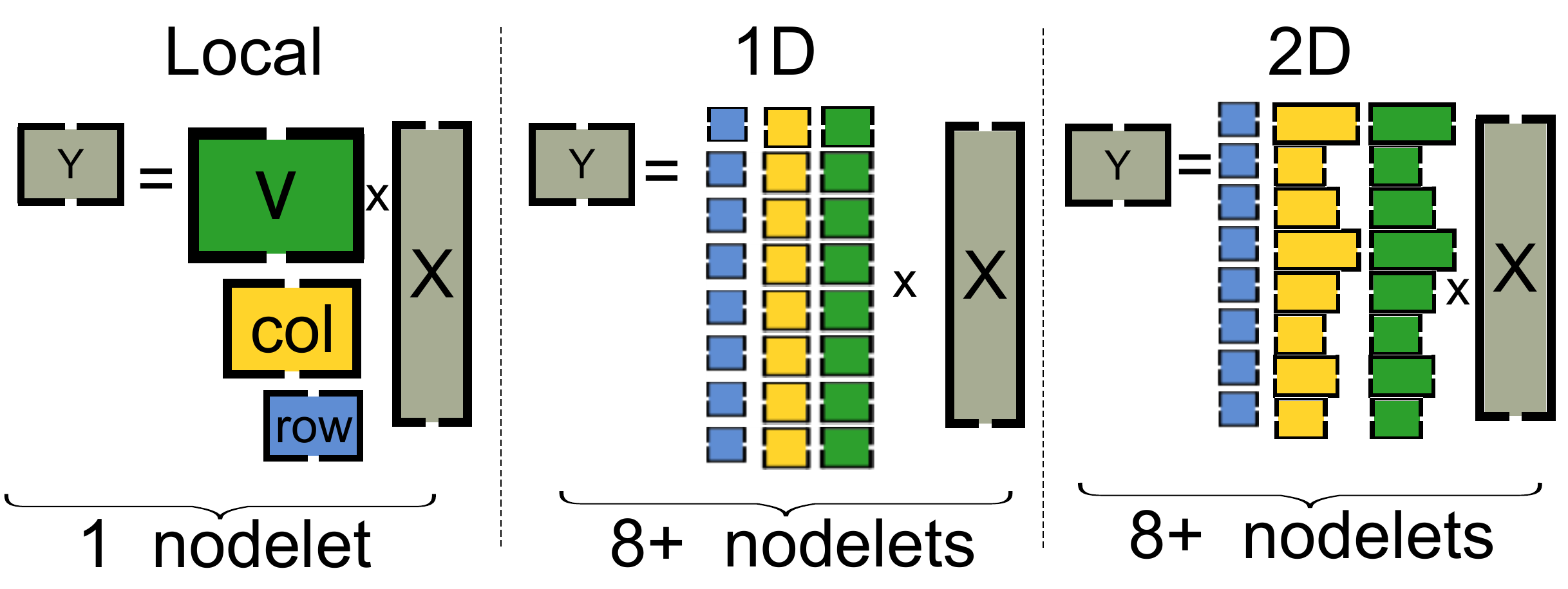}
  \caption{Emu-specific layout for CSR-SpMV}
  \label{fig:csr_spmv_layout}
\end{figure}

\paragraph{Sparse Matrix-Vector Multiplication (SpMV)} In addition to being a fundamental kernel for graph analytics and sparse tensor decomposition applications, SpMV provides an opportunity to investigate data layout strategies on the Emu's global physical address space. Emu provides a ``local'' malloc (\texttt{mw\_localmalloc}) similar to a traditional contiguous malloc as well as a ``striped'' malloc (\texttt{mw\_malloc1dlong}, called 1D) that places data in a round-robin fashion across nodelets and a two-dimensional malloc (\texttt{mw\_malloc2d}, called 2D), that distributes entire data structures across nodelets. 

Figure \ref{fig:csr_spmv_layout} demonstrates the three layouts that are tested with inputs in the popular Compressed Sparse Row (CSR) format. For an $I \times J$ matrix A with $M$ non-zeros, a size-$(I+1)$, a size-$M$, and another size-$M$ arrays are stored for row pointers, column indices, and nonzero values respectively. We use $y=Ax$ to illustrate an SpMV operation and use one node case as an example which can be extended to the multi-node case.
In the local case, contiguous mallocs are used to place all data on a single nodelet, which include the output vector \texttt{Y}, the input vector \texttt{X}, and the input CSR matrix: \texttt{row}, \texttt{col}, and \texttt{V}. Only one nodelet's computing power is employed.
1D layout stripes the \texttt{row, col}, and \texttt{V} arrays of the input matrix across the nodelets using \texttt{mw\_malloc1dlong} while \texttt{X} is replicated across all nodelets and \texttt{Y} is on nodelet 0. Due to the diverse sizes of \texttt{row}, \texttt{col}, and \texttt{V} and the round-robin pattern of \texttt{mw\_malloc1dlong}, the nonzeros and column indices in the same row are distributed to different nodelets, which introduces frequent thread migration.
For the 2D allocation, we use a two-stage allocation rather than Emu's 2D malloc function to partition \texttt{V} and \texttt{col} across multiple nodelets. First, the length of each row that is assigned to a nodelet is computed and then \texttt{V} and \texttt{col} are allocated in units of rows on each nodelet. We still employ the \texttt{mw\_malloc1dlong} function but use it to allocate a variable length array. \texttt{X} is also replicated across nodelets as is the output, Y. Using this modified 2D layout, we can wipe out most thread migrations with the exception for some initial spawn migrations and reduction at the end of the computation. This is an ideal case to test the highest performance Emu can achieve in the SpMV benchmark.



%% file: sections/results.tex
\section{Results} \label{sec:results}

The updated characterization of the Emu Chick repeats the STREAM, pointer chasing, and SpMV experiments that were initially investigated in \cite{hein:2018:ashes_emu}, but these experiments focus on further characterizing the entire Chick ``multi-node'' system that uses all 64 nodelets (across 8 nodes) within the Chick. Single node results are presented for specific experiments including SpMV layout and simulation, primarily due to limitations in the current firmware (i.e., certain configurations encounter hardware faults) and slowness of the Emu architectural simulator. 

\subsection{STREAM}
\label{sec:stream}

\begin{figure*}
  \centering
  \includegraphics[width=0.5\linewidth]{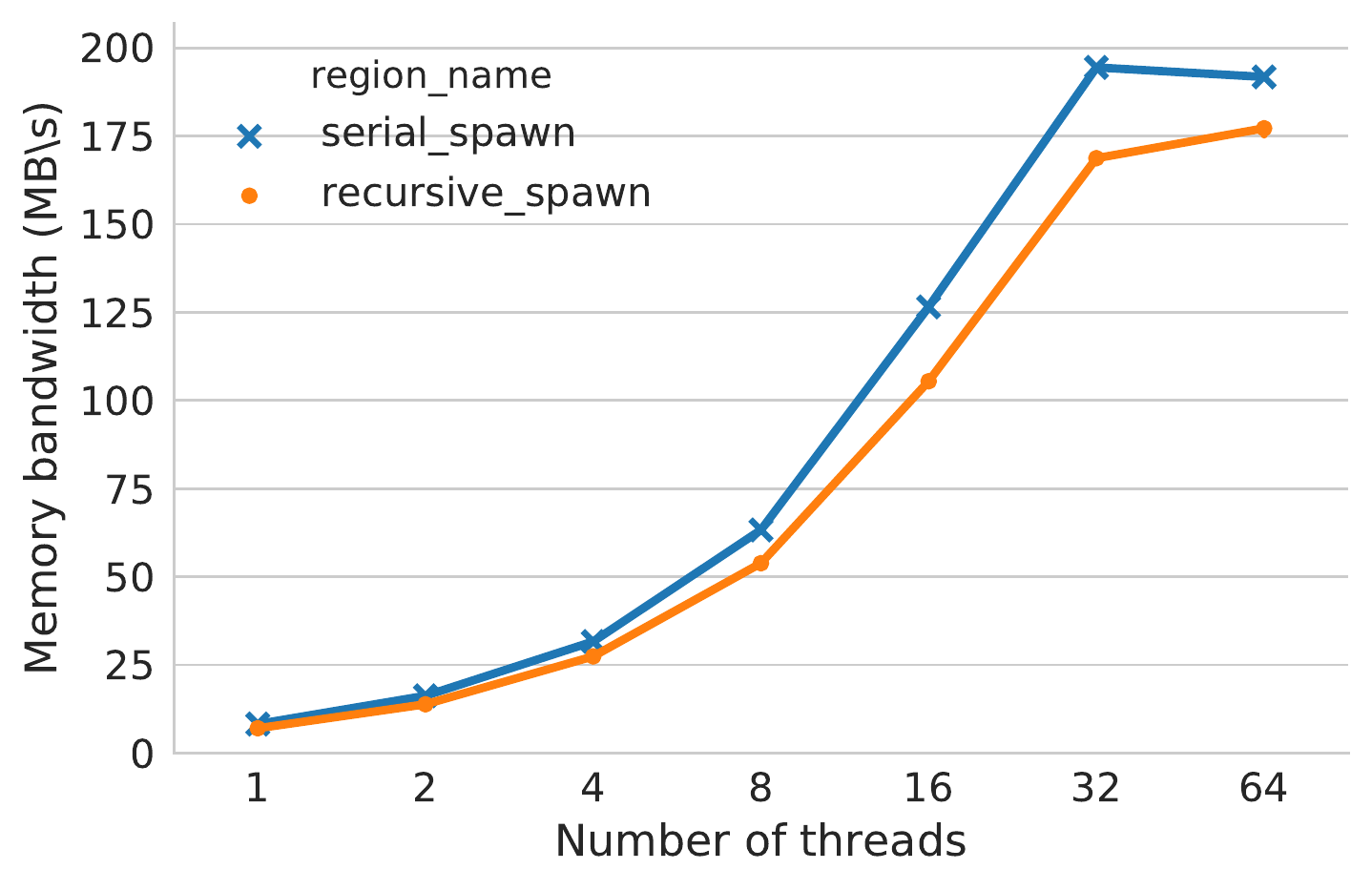}
   \caption{Memory bandwidth achieved on a single nodelet of the Emu Chick. Threads are created using a serial loop or a recursive spawn tree.}
  \label{fig:emu_local_stream}
\end{figure*}

Figure~\ref{fig:emu_local_stream} shows the results from running the STREAM benchmark on a single Emu nodelet. Performance scales up with thread count through 32 threads and then plateaus. Two methods of thread creation are tested here. In the \texttt{serial\_spawn} strategy, a single thread uses a for loop to create each worker thread, while \texttt{recursive\_spawn} uses a recursive spawn tree. There is not much difference between the two approaches, indicating that thread creation is not terribly expensive on the Emu platform.

\begin{figure*}
  \centering
  \begin{subfigure}[Local STREAM]{0.45\textwidth}
  \includegraphics[width=\textwidth]{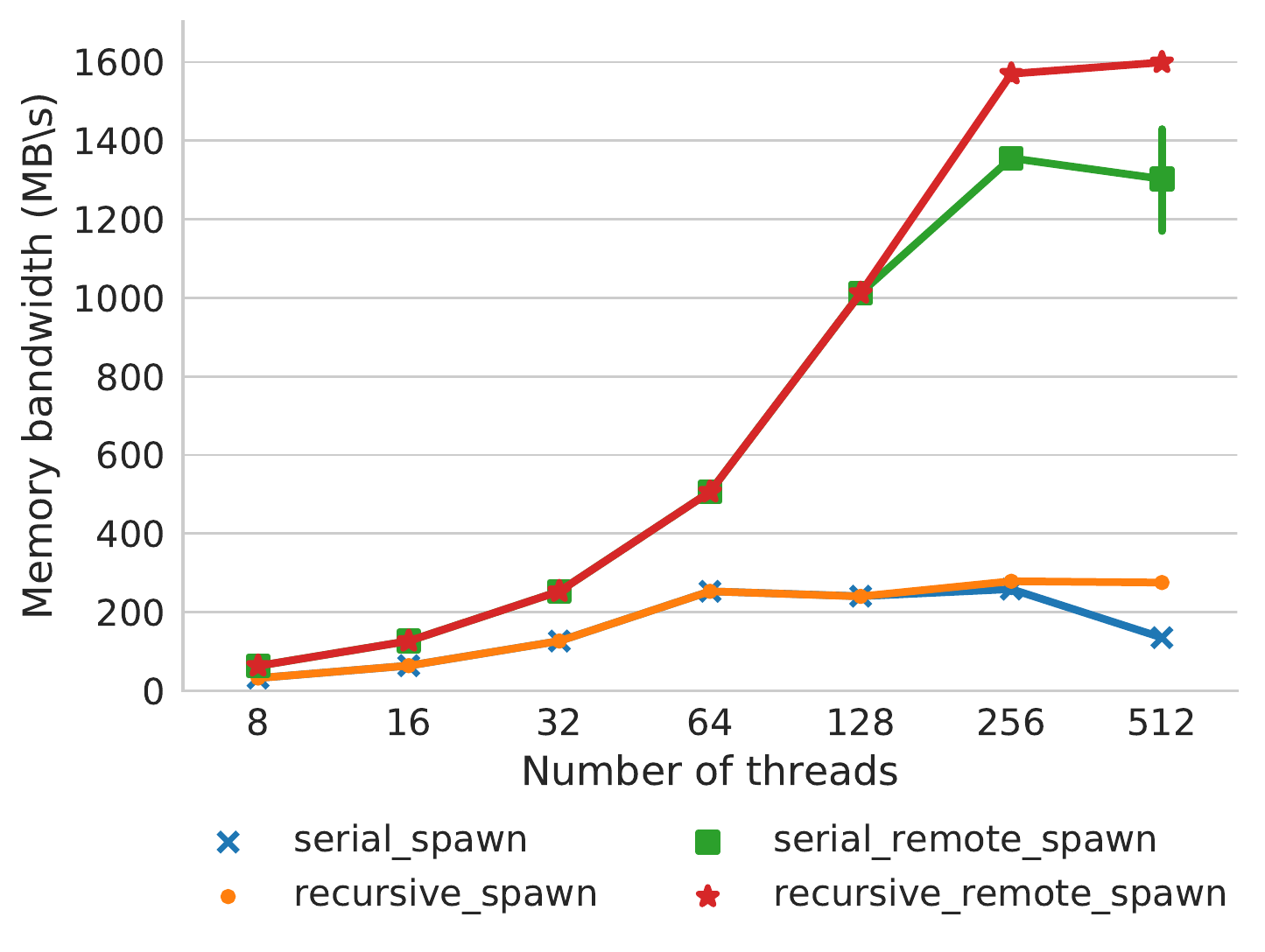}
\caption{Single Node (8 nodelets)}
\label{fig:emu_singlenode_global_stream}
\end{subfigure}
\begin{subfigure}[Local STREAM]{0.45\textwidth}
  \includegraphics[width=\textwidth]{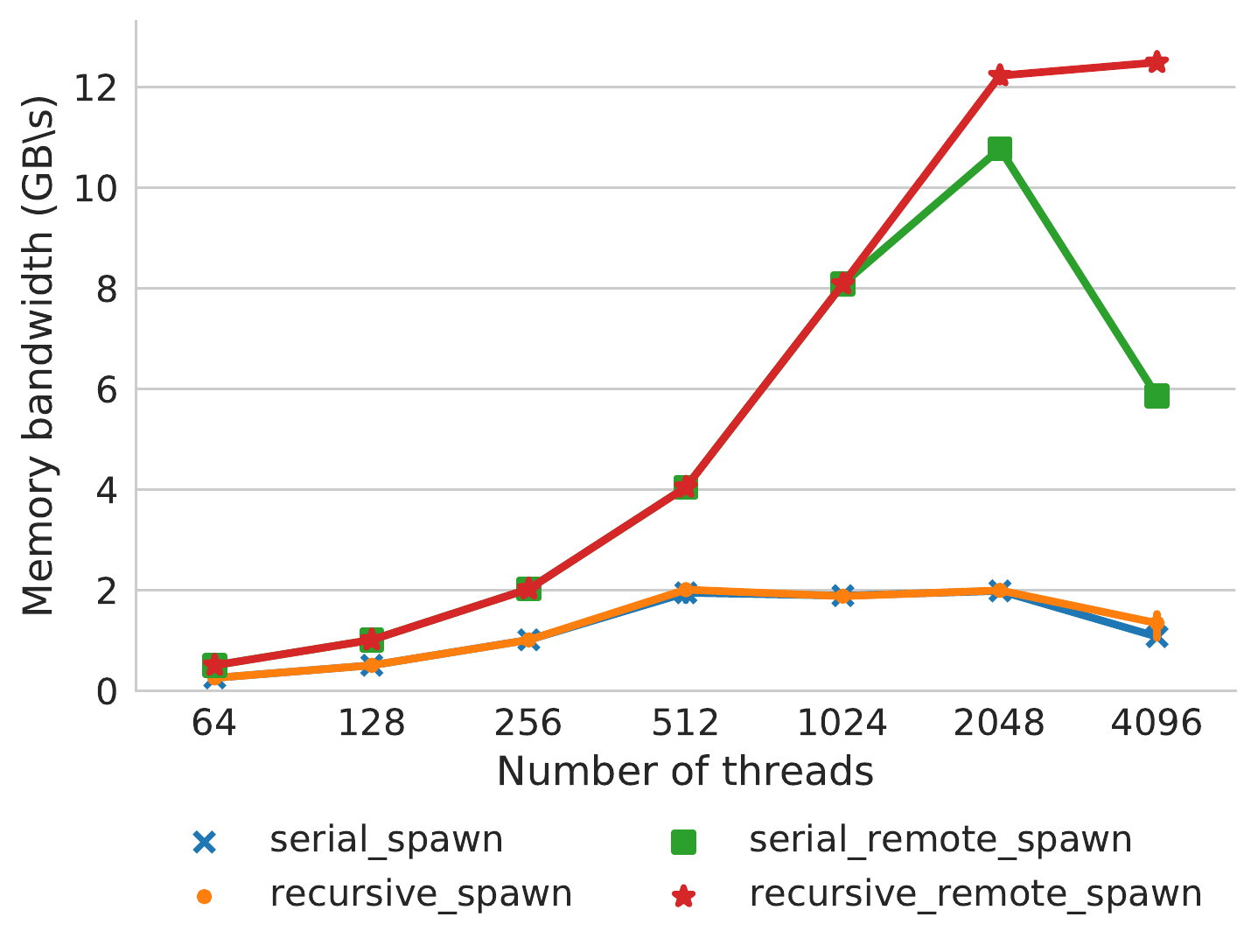}
  \caption{Multi-node (64 nodelets)}
 \label{fig:emu_global_stream}
 \end{subfigure}

\caption{Emu STREAM performance with different spawn strategies}
   \label{fig:emu_stream_spawn}
\end{figure*}

In Figure~\ref{fig:emu_singlenode_global_stream}, we extend the STREAM benchmark to run on eight nodelets (one node card) of the Emu Chick. Two new thread creation strategies are introduced here, \texttt{serial\_remote} \texttt{\_spawn} and \texttt{recursive\_remote\_spawn}. A remote spawn on Emu means that the thread is created on a remote nodelet, rather than being created locally and allowed to migrate to the remote data. The ``remote'' thread creation strategies first create a thread on each nodelet (either one at a time or with a recursive spawn tree), and then perform a second level of spawning on the local nodelet, as in the single nodelet case. 
Figure~\ref{fig:emu_global_stream} extends the analysis to 64 nodelets and up to 4096 threads showing that recursive remote spawn continues to scale for large numbers of threads up to 12 GB/s across 8 nodes. 
Both sets of results show that remote spawns are essential to achieving maximum bandwidth on Emu. 

In comparison to the Emu, the Xeon system (\texttt{Haswell}) achieves 100 GB/s on the STREAM benchmark (with an interleaved NUMA layout across four sockets) while the Emu Chick has a maximum STREAM bandwidth of 12 GB/s. The Emu bandwidth is currently limited by CPU speed and thread count rather than DDR bus transfer rates. However even with this prototype system we can observe improvements in other benchmarks where the memory access pattern is not as linear and predictable as it is with STREAM.

\subsection{Pointer Chasing}\label{sec:pointer-chasing}

\begin{figure*}
  \centering
  \includegraphics[width=0.8\linewidth]{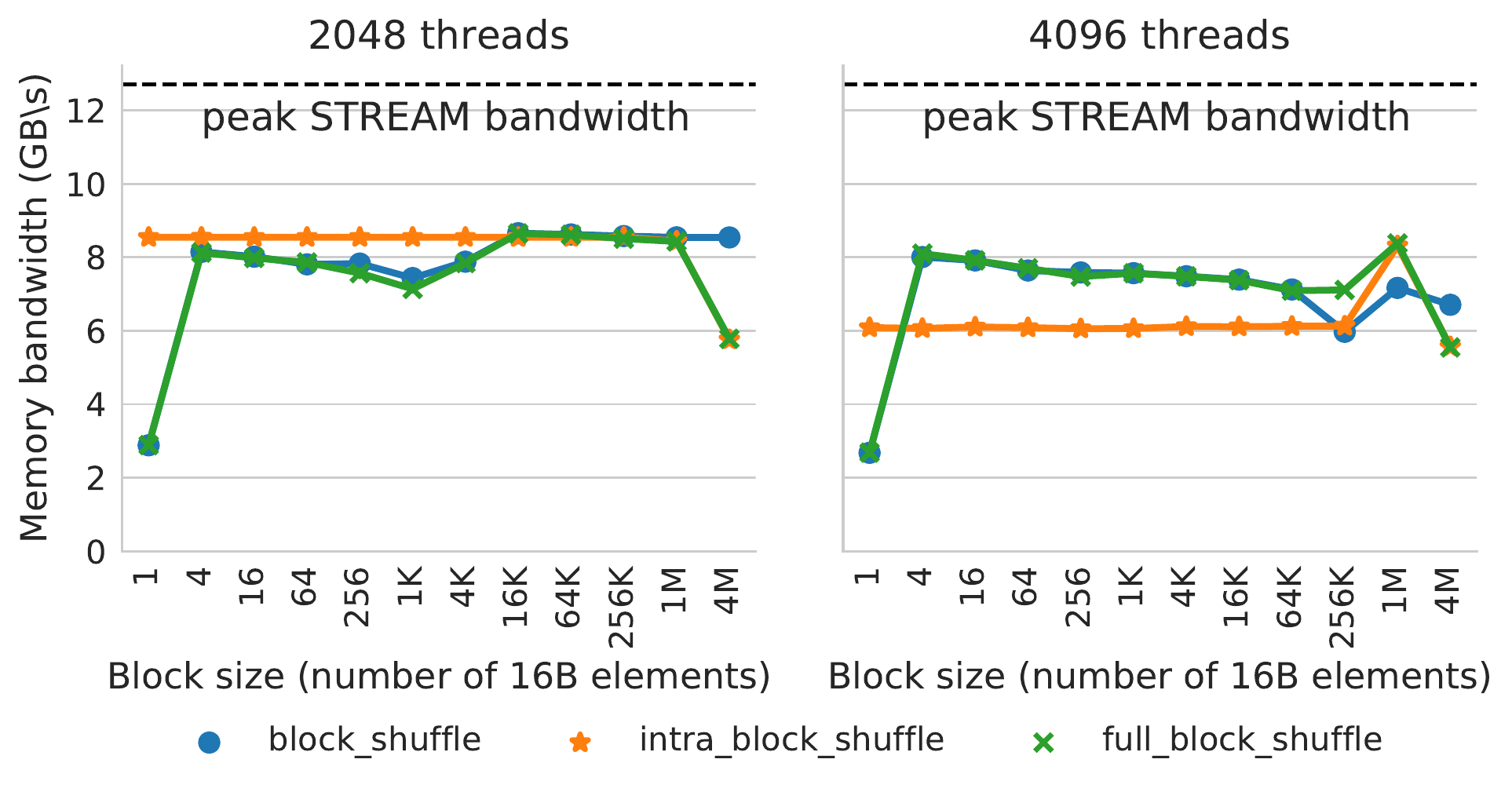}
  \caption{Pointer chasing performance on the Emu Chick (8 nodes, 64 nodelets). }
  \label{fig:emu_ptr}
\end{figure*}

\begin{figure*}
  \centering
  \includegraphics[width=0.8\linewidth]{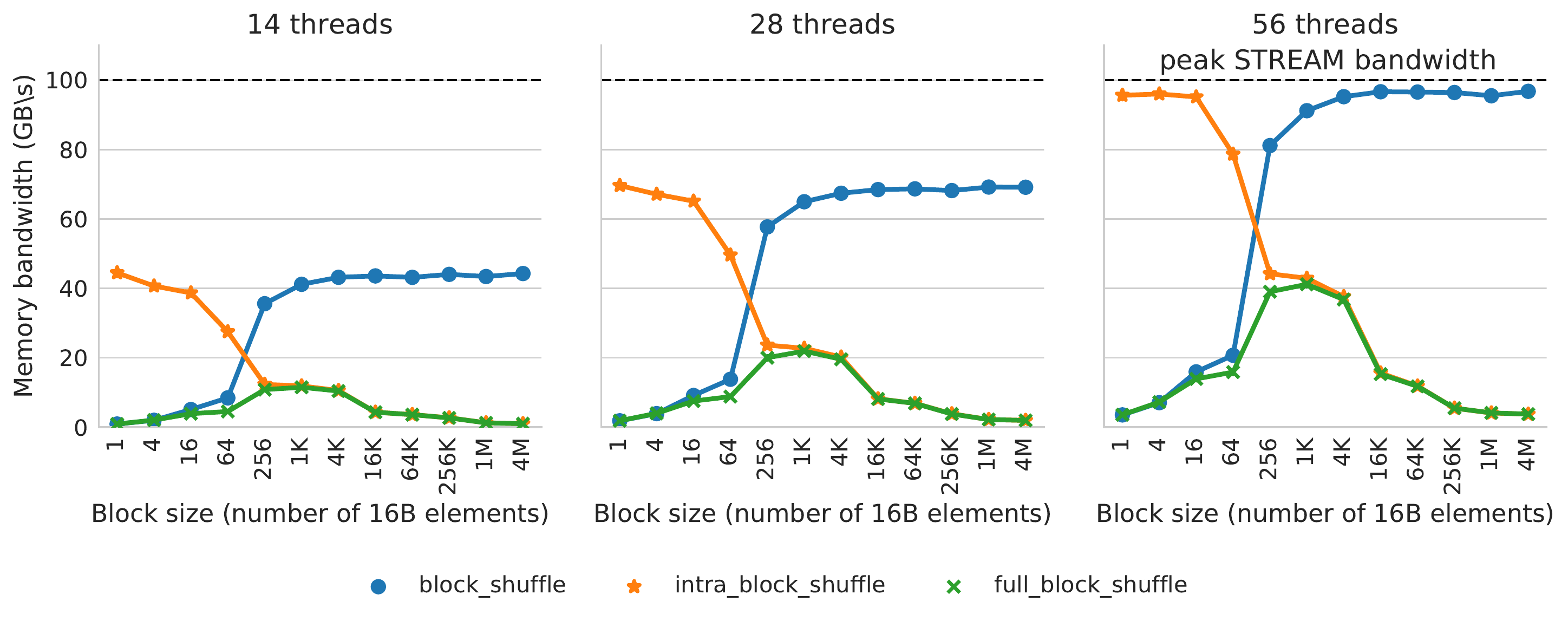}
  \caption{Pointer chasing performance on Haswell Xeon.}
  \label{fig:xeon_ptr}
\end{figure*}

Figures~\ref{fig:emu_ptr} and~\ref{fig:xeon_ptr} compare the performance of the Emu Chick against our Haswell Xeon server system for the pointer chasing benchmark. These results reveal important characteristics of both systems and highlight the unique advantages of the Emu Chick.

Pointer chasing on the Xeon architecture performs poorly for several reasons. For small block sizes, the memory system bandwidth is used inefficiently. An entire 64-byte cache line must be transferred from memory, but only 16 bytes will be used. The best performance is achieved with a block size between 256 and 4096 elements. This corresponds to a memory chunk of about 8KiB, the size of one DRAM page. Regardless of the size of the access, an entire DRAM row must be activated for each element traversed. Adding more threads at this point increases the number of simultaneous row activations. As the block size grows beyond the size of a DRAM page, performance declines again. 

Performance on Emu remains mostly flat regardless of block size. Emu's memory access granularity is 8 bytes, so it never transfers unused data in this benchmark. As long as a block fits within a single nodelet's local memory channel, there is no penalty for random access within the block. However, block size of 1 provides an interesting case; here Emu threads are likely to migrate on every access, and so performance is greatly reduced. But performance recovers when even as few as four elements are accessed between each migration.

\begin{figure*}
  \centering
  \includegraphics[width=0.6\linewidth]{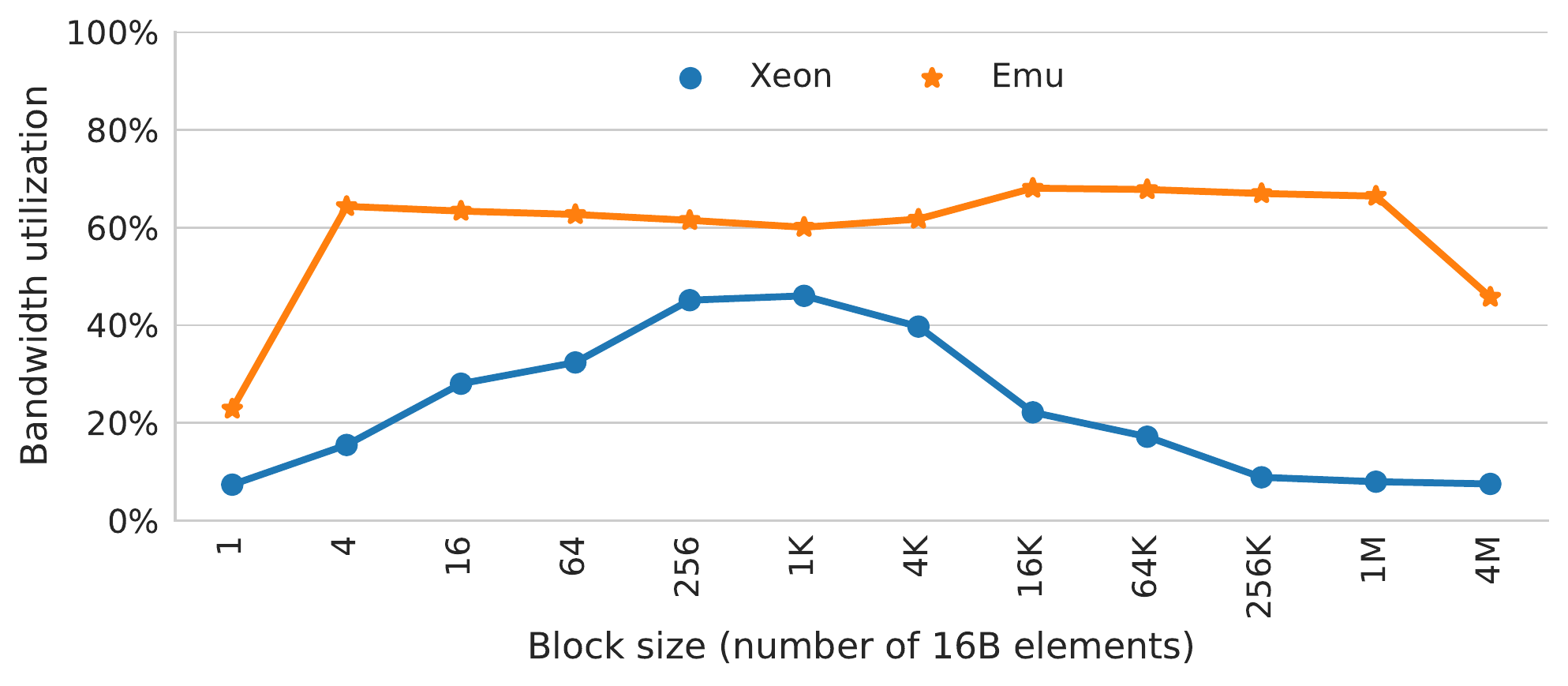}
  \caption{Bandwidth utilization of pointer chasing, compared between Sandy Bridge Xeon and Emu (64 nodelets)}
  \label{fig:xeon_vs_emu_ptr}
\end{figure*}

Figure~\ref{fig:xeon_vs_emu_ptr} shows the normalized bandwidth usage (\textit{i.e.}, effective bandwidth usage) for the Haswell and Emu systems. The performance of each system has been normalized to the peak measured bandwidth of the system (\textit{i.e.}, the best result on the STREAM benchmark). In the pointer chasing benchmark, the Emu system is much better at using the available system bandwidth, using ~65\% of available system bandwidth in most cases and 25\% in the worst case. The Haswell Xeon uses less than 50\% of peak bandwidth in most cases and less than 10\% in the worst case, relying on multi-kilobyte levels of locality to efficiently transfer the data. These results bode well both for the targeted streaming graph and tensor decomposition applications which have pointer chasing behavior and rely on random accesses to compute SpMV and SpMM (sparse matrix-matrix product) operations, respectively.

\subsection{Sparse Matrix-Vector Multiplication}
\label{sec:spmv_results}

\begin{figure*}
  \centering
    \includegraphics[width=0.5\textwidth]{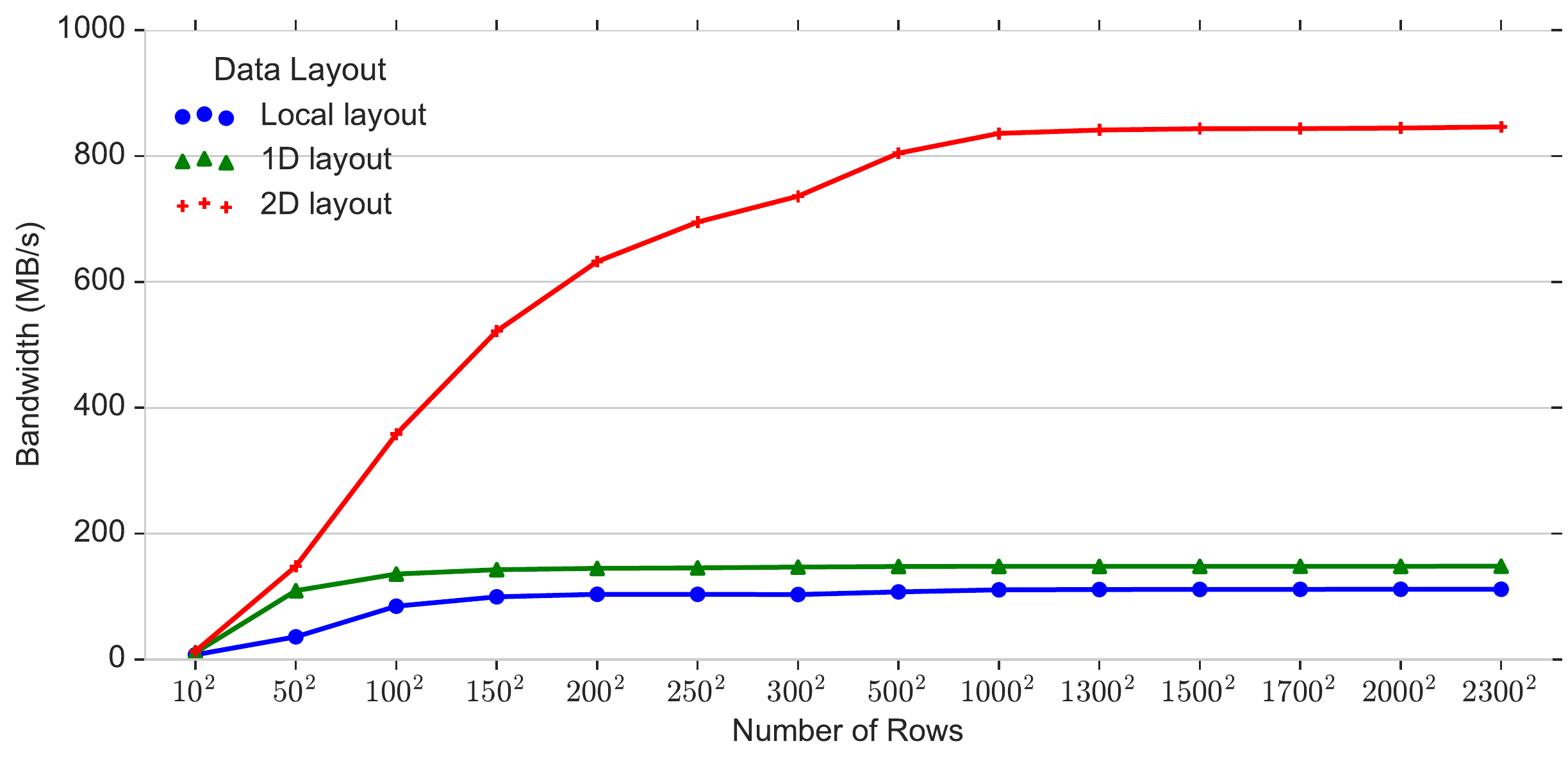}
  \caption{Bandwidth utilization of Emu Single Node (8 nodelets) with different SpMV Data Layouts}
  \label{fig:emu_spmv_dl_bw}
\end{figure*}

\begin{figure*}
\centering
\begin{subfigure}[x86 SpMV ]{0.44\textwidth}
  \includegraphics[width=\textwidth]{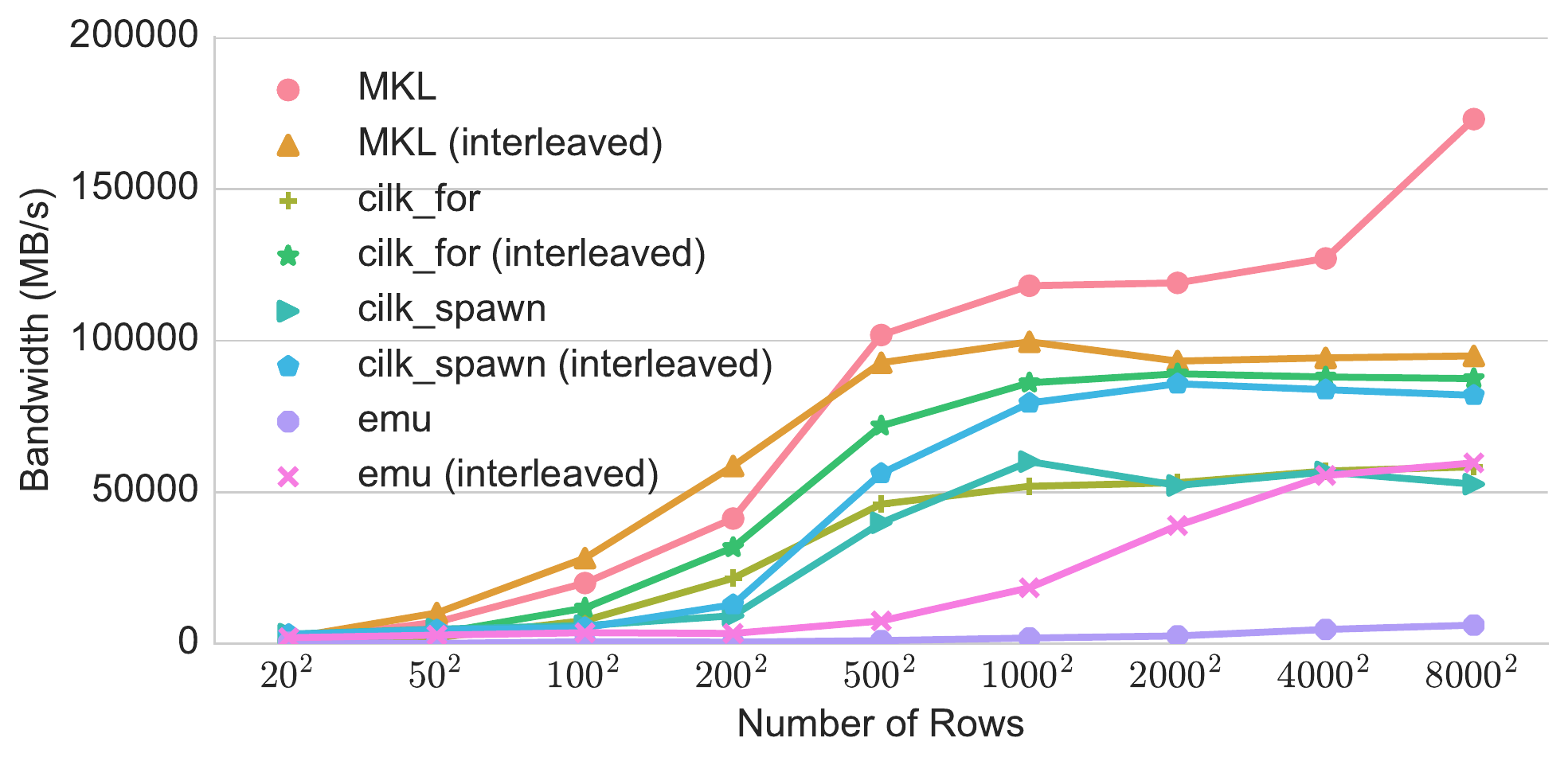}
\caption{Effective bandwidth for the Haswell Xeon}
\label{fig:cpu_spmv_bw}
\end{subfigure}
\begin{subfigure}[x86 SpMV ]{0.44\textwidth}
  \includegraphics[width=\textwidth]{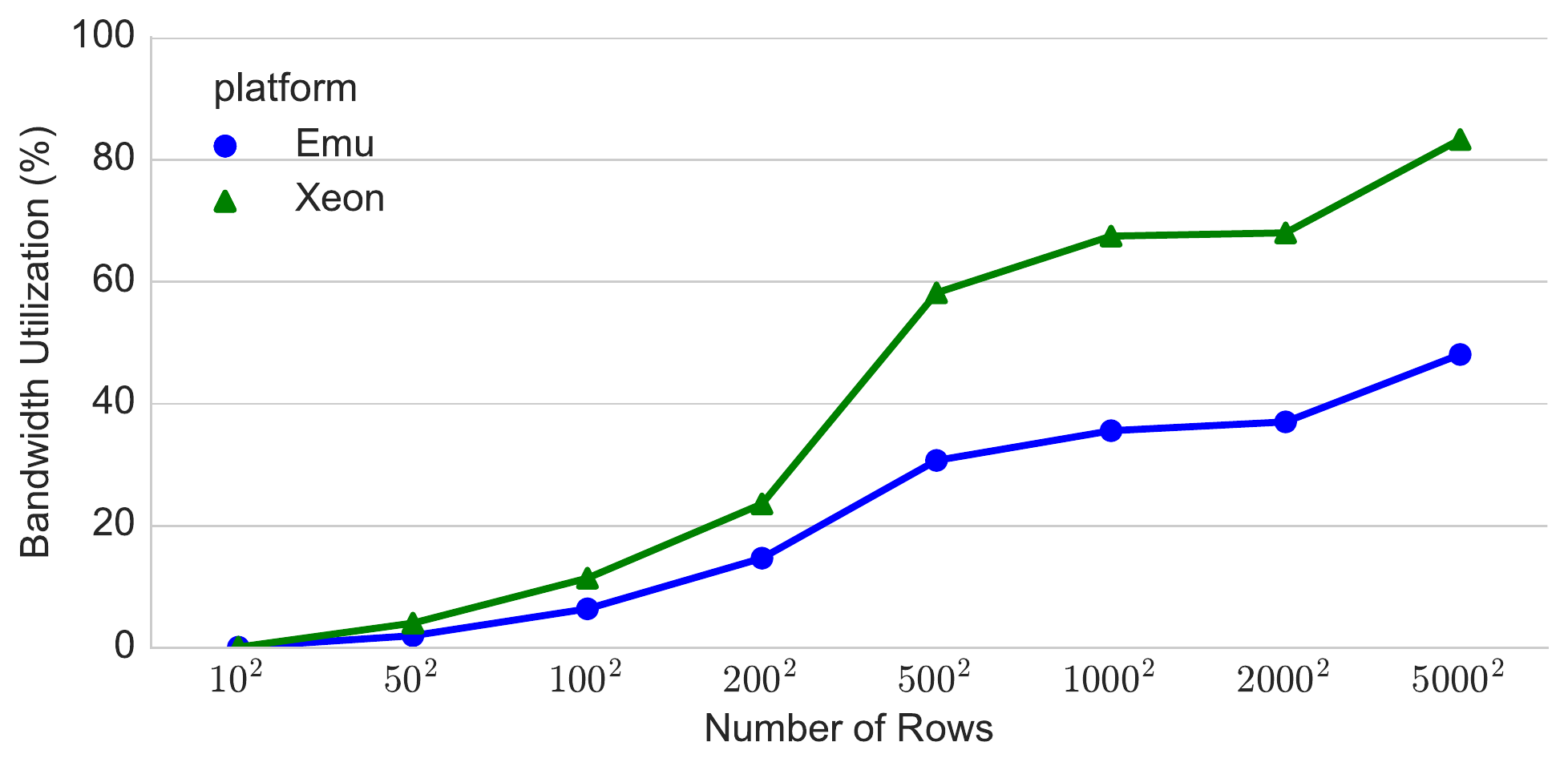}
  \caption{SpMV Percentage of STREAM}
  \label{fig:cpu_emu_spmv_pct}
\end{subfigure}
\caption{SpMV Emu Comparisons with Haswell Xeon}
  \label{fig:cpu_emu_spmv}
\end{figure*}

We use a synthetic Laplacian matrix as an input corresponding to a $d$-dimensional $k$-point stencil on a grid of length $n$ in each dimension. $d=2$ and $k=5$ specify a sparse matrix generated from a 5-point, 2-D, $n \times n$ stencil. With the Laplacian size $n$, the matrix is $n^2 \times n^2$  with 5 diagonals.
CPU tests are run on the Haswell Xeon-based system described in Section~\ref{sec:cpu-specs}, using SpMV from Intel's Math Kernel Library (MKL) with \texttt{MKL\_MAX\_THREADS} set at 56 (the number of physical cores in the system as opposed to total threads). We include two Cilk SpMV kernels for comparison, labeled \texttt{cilk\_for} and \texttt{cilk\_spawn}, which are written with the respective Cilk primitives, compiled using GCC 5.5.0, and run with \texttt{CILK\_NWORKERS} set to 56. Data is distributed across NUMA regions using \texttt{numactl --interleave=0-3}. 

Figure~\ref{fig:emu_spmv_dl_bw} shows the memory bandwidth achieved by SpMV on a single node (8 nodelets) of the Chick using each of the three data layout strategies: local, 1D, and 2D layouts. The local layout on the Emu suffers from a limited amount of thread parallelism while the 1D layout suffers from a large number of thread migrations, resulting in max bandwidths of close to 96.13 MB/s and 148.66 MB/s, respectively. Overall, the 2D memory layout provides the only scalable solution for SpMV, scaling up to 846.39 MB/s for n=2300. 

Figure~\ref{fig:cpu_spmv_bw} shows results for experiments run on the Haswell Xeon machine with four different code configurations and two NUMA layouts, the default ``current" policy (place all data in current node) and ``interleaved" which interleaves data across four sockets at a page granularity. 1) \textit{MKL} refers to an implementation of SpMV using Intel's MKL library, 2) \textit{cilk\_for} is a native Cilk implementation using cilk\_for, 3) \textit{cilk\_spawn} is a native Cilk implementation using cilk\_spawn, and 4) \textit{emu} is an implementation of SpMV that is backported from the Emu-optimized version of the code. This last version of the code includes the 2D layout optimization described in Section \ref{ssec:bmks} and evaluated in Figure \ref{fig:emu_spmv_dl_bw}. 

The Haswell Xeon results in Figure \ref{fig:cpu_spmv_bw} show that the MKL implementation of SpMV with the ``current" NUMA layout achieves the highest bandwidth, getting close to 175 GB/s across four nodes. Meanwhile, \textit{cilk\_for} and \textit{cilk\_spawn} show similar scaling up to n=200 with the interleaved versions of both implementations closely mirroring each other's performance from n=500 to n=8000. Finally, the Emu ``backported" code only shows scalable performance with the NUMA interleaved setting, peaking in performance at n=8000 and around 50 GB/s. While it is unclear at the moment why the MKL version scales so well with a non-interleaved data layout (we suspect it may have to do with first-touch layouts being amenable to relevant MKL data structures and computation), the performance of the Emu backported code seems to mesh well with our understanding of the Emu system as a distributed PGAS machine. That is, the optimal 2D layout of SpMV for the Emu keeps data accesses ``local" on the Emu, but a normal x86 NUMA allocator does not stripe data like Emu's allocator does, meaning that most data accesses on the x86 emu ``current" setup are remote NUMA accesses. Furthermore, x86 NUMA interleaved layouts have much larger granularity for striping (pages versus elements in an array on the Emu), which likely also penalizes the ``emu (interleaved)" implementation. 

Following the Haswell Xeon results, we compare the total percentage of STREAM bandwidth that is achieved for SpMV on the Emu versus on the Haswell machine in Figure \ref{fig:cpu_emu_spmv_pct}. The Emu results use the peak multi-node STREAM bandwidth of 12 GB/s and are compared to the Haswell STREAM peak \textit{without NUMA interleaving}, which peaks at 175 GB/s. In the case of the Haswell results, the best case SpMV (MKL non-interleaved from Figure\ref{fig:cpu_spmv_bw}) is used as the comparison point. As opposed to the pointer chasing results in Figure \ref{fig:xeon_vs_emu_ptr}, we see that both systems scale in terms of bandwidth utilization as the amount of synthetic data is increased with the Emu peaking at about 50\% of peak STREAM bandwidth versus the Haswell system's 80\% of peak STREAM bandwidth. SpMV can achieve between 50\% to 60\% of the peak STREAM bandwidth, but with its address calculations and the multiply-and-accumulate, the 175MHz Gossamer Cores cannot generate loads quickly enough to saturate the available memory bandwidth. As we discuss in the following section, a primary limitation of the current Chick prototype is that even simple benchmarks are compute-bound.

%% file: sections/discussion.tex
\section{Discussion} \label{sec:disc}

This characterization raises important topics for programming memory-centric architectures like the Emu Chick and also for building realistic comparisons between prototype novel architectures and existing architectures.

\subsection{Achievable Bandwidth for the Current Emu Chick Prototype}
\label{sec:stream_theory}

Using STREAM, pointer chasing, and SpMV, we provide an initial look at the performance of the Emu Chick for these fundamental operations. However, these results also point to a fundamental issue with the initial Emu Chick prototype. The design of the Chick with 1 Gossamer Core per nodelet and 8 nodelets per node leads to a situation where most codes are currently compute bound due to limited GCs and low frequencies of the GCs on the FPGA prototype. 

To demonstrate, we look at the inner loop of a STREAM ADD operation in Listing~\ref{lst:stream-add} and analyze the relevant assembly code generated by Emu's gossamer64-objdump tool. As Listing \ref{lst:stream-add-asm} shows, each inner loop of the STREAM ADD kernel performs two loads, one add operation, and one store operation for a total of 3 memory operations  out of a total of 21 instructions. 

\noindent\begin{minipage}{.48\textwidth}
\begin{lstlisting}[
  language=C,frame=tlrb,basicstyle=\ttfamily\scriptsize,
  label=lst:stream-add,
  caption={STREAM ADD worker function}
]
noinline void
recursive_remote_spawn_level2_worker(long begin, 
long end, long * a, long * b, long * c)
{
    for (long i = begin; i < end; ++i) {
        c[i] = a[i] + b[i];
    }
}      
\end{lstlisting}
\end{minipage}

\begin{minipage}{.4\textwidth}
\begin{lstlisting}[
  language=C,frame=tlrb,basicstyle=\ttfamily\scriptsize,
  label=lst:stream-add-asm,
  caption={STREAM ADD Assembly}
]
%for.body:             // block
	etd       2    // D = E2
	sllc      3    // D <<= 3
	dpeta     6    // A = D + E6
	lde       7    // LOAD: E7 = *A
	etd       2    // D = E2
	sllc      3    // D <<= 3
	dpeta     5    // A = D + E5
	lde       8    // LOAD: E8 = *A
	etd       2    // D = E2
	sllc      3    // D <<= 3
	dpeta     4    // A = D + E4
	etd       8    // D = E8
	adde      7    // ADD: D += E7
	wrd            // STORE: *A = D
	eta       2    // A = E2
	aaimb     1    // A += 1
	ate       2    // E2 = A
	etd       3    // D = E3
	xore      2    // D ^= E2
	bdz       %for.end
	jmp       %for.body
%for.end:              // block
\end{lstlisting}
\end{minipage}

Using this information from the assembly code, we can determine the peak achievable bandwidth by one GC running at the current frequency of 175 MHz. As Equation \ref{eq:stream-add} shows, one GC can achieve up to 200 MB/s for the STREAM ADD kernel, which we have used as our "peak" achievable bandwidth for comparison with other microbenchmarks.  

\begin{equation}
\label{eq:stream-add}
\begin{split}
175MHz \Rightarrow \frac{\text{175M cycles}}{second} \times \frac{\text{1 instruction}}{cycle} \times\\ \frac{\text{3 mem ops}}{\text{21 instructions}} \times \frac{\text{8 Bytes}}{\text{1 mem op}} = 200 MB/s
\end{split}
\end{equation}

Investigating further in Table \ref{tbl:stream}, we see that the measured results for STREAM ADD for single-node and multi-node execution are very close to the 200 MB/s peak value for STREAM ADD. However, if we look at the ideal case where all instructions are memory operations, the peak value for a single GC would be closer to 1,400 MB/s. Moreover, looking at the Chick's memory system design, we see that the measured results for STREAM ADD are 8x slower than the NCDIMM's theoretical peak achievable bandwidth.

This analysis points to one conclusion - the current Emu Chick prototype is compute-bound for all microbenchmarks due to a low number of GCs and by low frequencies of the GCs, both of which are restricted by limited FPGA area and speed grades in the Arria 10 host device. We estimate that 8 GCs at the same speed of 175 MHz per nodelet or NCDIMM channel would be needed to move from a regime where the Chick is compute-bound to one where applications are memory-bound. 

\begin{table}
\centering
\caption{STREAM and Memory Bandwidths (BW in MB/s)}
\label{tbl:stream}
\begin{tabular}{lrrrr}
\toprule
Operation &   \multicolumn{1}{c}{Nodelets} &   \multicolumn{1}{c}{Scale} &  \multicolumn{1}{c}{Threads} &  \multicolumn{1}{c}{BW} \\
\midrule
 ADD (Measured) &   8 &   30 & 512 & 1,600\\
 ADD (Measured) &   64 &  31 & 4096 &  12,790\\
 Ideal - all \textit{ld} ops & 1 &    & & 1400\\
 NCDIMM & 8 & & & 12,800\\
 NCDIMM & 64 & & & 102,400\\
\bottomrule
\end{tabular}
\end{table}

\subsection{Caveats for Programming the Emu Chick}
While Cilk provides an easy entry point for programming microbenchmarks for the Chick, our initial characterization has demonstrated some pitfalls for obtaining good performance on the Chick prototype. Primarily, the programmer must consciously design algorithms that optimize data layouts across nodes and that limit load imbalance by limiting thread migration. While CPU-based systems typically are optimized using techniques like cache-blocking, the distributed Partitioned Global Address Space (PGAS) nature of the Chick system requires that the programmer explicitly think about data placement in a different fashion. The Emu Chick has an existing profiler to inspect postmortem where threads end have migrated to, but detailed profiling and inspection of a program's execution requires the use of a simulation-based profiling tool.  

The results from SpMV demonstrate that data layout can have an impact on performance on the Emu, application performance also depends on where threads are spawned and how many migrations occur between nodes and nodelets.
In the initial development of our benchmarks, we debated explicitly minimizing thread movement and keeping computation local to a specific node to limit load imbalance on the existing compute resources for each nodelet. However, this strategy both goes against the ``lightweight, migrating threadlets'' model of computation with the Emu, and it is hard to implement in practice. 

For this reason, we have settled on a strategy of ``smart thread migration'' for future benchmarking and application development with the Emu system. In short, this means
\begin{inparaenum}[1)]
  \item using ``smart'' thread spawn techniques like the two-level recursive remote spawn as in Section~\ref{sec:stream},
  \item using replicated allocations for commonly used inputs like the vector \texttt{X} in the SpMV benchmark, and 
  \item picking the appropriate layout strategy for the application.
\end{inparaenum}
In this last case, it is likely that good application performance will be most easily achieved through proper data layouts like with CSR SpMV's striped allocation across nodelets and per-nodelet secondary allocation for different-length rows. In this sense, we have created our own custom 2D allocator for SpMV, but we expect that higher-level memory allocation constructs will eventually be supported to help use the Emu's novel global address space layout.  

\subsection{Performance Models and Comparisons to Existing Architectures}

One of the challenges in evaluating a drastically different architecture like the Emu is performing a realistic comparison between a prototype architecture and existing platforms using CPUs or other mainstream accelerators.
Many aspects of the prototype Emu Chick present challenges.
The Chick is a cacheless architecture and uses thread migration and atomic operations to avoid buffering large chunks of data.
Even when compared with accelerators like GPUs, the low-latency access of the Chick, different memory clock speeds and data widths, and the lack of shared memory or caches provide a challenge for modeling how much more ``efficient'' the Chick is in terms of memory bandwidth. As shown in Section \ref{sec:results}, different STREAM numbers for x86 systems based on NUMA interleaving settings also complicates this comparison. Additionally, the Chick is a full-scale prototype built using FPGA devices, which are useful for their flexibility and customization capabilities but naturally are slower than a traditional, hardwired ASIC.
Firmware upgrades to the Chick prototype can also affect application performance dramatically by changing the gossamer cores' maximum frequency and by adding new functionality.

These comparison challenges are common not only to the Emu Chick but also to other new, experimental hardware like neuromorphic and quantum computing platforms.
We may need to define additional metrics to supplement traditional characterization metrics like performance (FLOPS), memory bandwidth balance (FLOPS/B), and power efficiency (FLOPS/W). While we do not yet have enough application experience with the Emu Chick to fully define new metrics, we propose that there may be promise in focusing on comparison metrics that highlight the differences listed above. For example, a cache-less system like the Emu Chick may not actually move data physically across the system, but a comparable metric to a traditional CPU-based system might be some combination of network traffic (\textit{i.e.}, threads migrated measured using context size and time, or B/s) and cache misses avoided (B/s). We plan to investigate how to better model and define these types of differences in future work to effectively quantify not just the high-level application benefits of novel architectures like the Chick but also the fundamental qualities that help define which applications are the best fit for these new architectures.


%% file: sections/related-work.tex
\section{Related Work}
\label{sec:related-work}

Advances in memory and integration technologies provide opportunities for profitably moving computation closer to data\cite{Siegl:2016:DCF:2989081.2989087}.
Some proposed architectures return to the older processor-in-memory (PIM) and ``intelligent RAM''\cite{592312} ideas.
Simulations of architectures focusing on near-data processing\cite{7429299} including in-memory\cite{8013497} and near-memory\cite{7056040} show great promise for increasing performance while also drastically reducing energy usage.
Other than our previous study\cite{hein:2018:ashes_emu}, and related work on characterizing the Emu by other research groups\cite{belviranli:2018:emuhpec,minutoli2015implementing} few of these architectures have been implemented in hardware, even FPGAs, limiting the data scales on which applications can be evaluated.

Other hardware architectures have tackled massive-scale data analysis to differing degrees of success.
The Tera MTA / Cray XMT\cite{5161108,GraphCT-Wiley-Chap} could provide high bandwidth utilization by tolerating long memory latencies in applications that could produce enough threads.
In the XMT all memory interactions were remote incurred the full network latency.
The Chick instead moves threads to memory on reads, assuming there will be a cluster of reads for nearby data.  The Chick processor needs to tolerate less latency and need not keep as many threads in flight.
Also, unlike the XMT, the Chick runs the operating system on the stationary processors, currently PowerPC, so the Chick processors need not deal with I/O interrupts and highly sequential OS code.
Similarly to the XMT, programming the Chick requires language and library extensions.
Future work with performance portability frameworks like Kokkos\cite{CarterEdwards20143202} will explore how much must be exposed to programmers.
Another approach is to push memory-centric aspects to an accelerator like Sparc M7's data analytics accelerator\cite{7091786} for database operations or Graphicionado\cite{7783759} for graph analysis.

Moving computation to data via software has had a successful history in supercomputers and clusters via {Charm++}\cite{7013040}, which manages dynamic load balancing on distributed memory systems by migrating the computational objects.
Previously data analysis systems like Hadoop had moved computation to data when the network was a data bottleneck, but that no longer appears to be useful\cite{Ananthanarayanan:2011:DDC:1991596.1991613}.

Finally, algorithms research related to SpMV could prove beneficial to future implementations for Emu-like architectures. New state-of-the-art SpMV formats and algorithms such as SparseX, which uses the Compressed Sparse eXtended (CSX) format for storing matrices\cite{Elafrou:2018:sparsex} provide an alternative data structure and data layout that can be used to improve the performance of SpMV-based operations on the Emu. Related to our characterization, other researchers have investigated techniques \cite{rolinger:2018:spmv_emu_ia3} to implement SpMV with a focus on creating a load-balanced implementation using BFS and the METIS graph partitioner to place pieces of a graph on different nodelets. Note that this load balancing reduces thread migrations and hotspots but may require a large amount of initial preprocessing.

Other recent work has also looked to extend from low-level characterizations like those presented here by providing initial Emu-focused implementations of Breadth-First Search\cite{belviranli:2018:emuhpec}, 
Jaccard index computation 
\cite{krawezik:2018:jaccard_emu_hpec}, bitonic sort, \cite{velusamy:2018:sort_emu_hpec} and compiler optimizations like loop fusion, edge flipping, and remote updates to reduce migrations
\cite{chatarasi:2018:psc_emu_mchpc}.


%% file: sections/conclusion.tex
\section{Conclusion} \label{sec:concl}

Our microbenchmark evaluation of the Emu Chick demonstrates some of the limitations of the existing prototype system as well as some potential
benefits for massive data analytics applications like streaming graph analytics and sparse tensor decomposition.
We demonstrate multi-nodelet (64 nodelets across 8 nodes) performance for a variety of benchmarks
including STREAM, pointer chasing, and SpMV.
Initial results demonstrate relatively low overall bandwidth for the Emu system with a peak of 12 GB/s STREAM bandwidth for the current Chick prototype (compared to 80+ GB/s on a Haswell CPU server socket). However, we also show that algorithms implemented on the Emu can achieve a high percentage of effective memory
bandwidth even in a worst-case access scenario like pointer chasing.
The pointer chasing benchmark in Section~\ref{sec:pointer-chasing}
achieves a stable 60-65\% bandwidth utilization across a wide range of
locality parameters.
These pointer chasing results and data layout studies show how random accesses with SpMV can be improved and while performance of SpMV does not quite match a well-optimized x86 implementation, these optimizations can provide a template for future benchmarking and
application development and show how application memory layouts and
``smart'' thread migration can be used to maximize performance on the
Emu system.
